\documentclass[usenatbib]{mn2e}
\usepackage{amsmath}
\usepackage{graphicx, color}
\usepackage{amssymb}
\newcommand{\avR}{\langle R \rangle}
\newcommand{\avT}{\langle T \rangle}
\newcommand{\Vsur}{V_{\rm survey}}
\newcommand{\DA}{D(z)}
\newcommand{\hz}{H(z)}

\newcommand{\Veff}{V_{\rm eff}}
\newcommand{\kmax}{k_{\rm max}}

\newcommand{\kmin}{k_{\rm min}}

\makeatother

\makeatother

\begin{document}
\setlength{\unitlength}{1mm} %\twocolumn[\hsize\textwidth\columnwidth\hsize\csname@twocolumnfalse\endcsname]

\title[Robustness to systematics]{Robustness to systematics for future dark energy probes}

\author[March et al]{Marisa C. March$^{1}$, Roberto Trotta$^{1}$, Luca Amendola$^{2}$ and Dragan Huterer$^{3}$\\
$^1$Astrophysics Group, Imperial College London, Blackett Laboratory,
  Prince Consort Road, London SW7 2AZ, UK \\ 
$^2$ITP, University of Heidelberg, Heidelberg, Germany \\  
$^3$Department of Physics, University of Michigan, 450 Church St., Ann Arbor, MI 48109 }

\maketitle

\begin{abstract}
We extend the Figure of Merit formalism usually adopted to quantify the
statistical performance of future dark energy probes to assess the robustness
of a future mission to plausible systematic bias. We introduce a new
robustness Figure of Merit which can be computed in the Fisher Matrix
formalism given arbitrary systematic biases in the observable quantities. We
argue that robustness to systematics is an important new quantity that should
be taken into account when optimizing future surveys. We illustrate
our formalism with toy examples, and apply it to future type Ia supernova
(SNIa) and baryonic acoustic oscillation (BAO) surveys. For the
simplified systematic biases that we consider, we find that 
SNIa are a 
somewhat more robust probe of dark energy parameters than the BAO. We trace
this back to a geometrical alignement of systematic bias direction with
statistical degeneracy directions in the dark energy parameter
space. 
\end{abstract}
\maketitle


\section{Introduction}

%\dragan{The pages are cut off at bottom when printed on US Letter paper (this
%  happens often with MNRAS-stype publications!) I know only how to fix it
%  using {\tt topmargin -1.5cm} command. Anyone know a better way with which to
%  send it to arXiv?}

%\dragan{Reworded intro a bit.}

%% The accelerated expansion of the Universe in relatively recent cosmic times
%% has been attributed to the existence of dark energy, whose characteristic is a
%% negative equation of state parameter, $w = p/\rho$, where $p$ is the fluid's
%% pressure and $\rho$ its energy density. Many models for dark energy have been
%% put forward, and one of the ways of distinguishing among them is to measure
%% possible departures in the equation of state from what is expected for a
%% cosmological constant, namely $w=-1$ at all redshifts.

The discovery of the accelerating universe has been hailed as one of the most
important developments in cosmology in decades. Yet the physical nature of the
cause of this acceleration -- the dark energy -- is lacking. With that
in mind, a large amount of effort has gone into measuring the parameters
describing dark energy, notably its equation of state parameter, $w = p/\rho$,
where $p$ and $\rho$ are the dark energy pressure and energy density. In fact,
measuring $w$ to high accuracy is one of the most important goals of ongoing
and upcoming large scale cosmological surveys, such as Dark Energy Survey
(DES\footnote{http://www.darkenergysurvey.org}), Baryon Oscillation
Spectroscopic Survey (BOSS\footnote{http://cosmology.lbl.gov/BOSS/}), Large
Synoptic Survey Telescope (LSST\footnote{http://www.lsst.org}), Square
Kilometre Array (SKA\footnote{http://www.skatelescope.org/}), Euclid
\citep{2009arXiv0912.0914L}.

In order to rank proposed future dark energy missions according to their
potential capabilities, a series of Figures of Merit (FoMs) has been
introduced, whose aim is to quantify the science return of an experiment in
terms of its ability to constrain dark energy.  Perhaps the most widely used
FoM is the one identified by the dark energy task force
(DETF)~\citep{Albrecht:2006um,Albrecht:2009ct}, which is a measure of the
statistical power of a future dark energy mission.  Other higher dimensional
versions have also been considered
\citep{HutererTurner,Albrecht_Bernstein,Wang_FoM,Crittenden_Pogosian_Zhao,Mortonson:2010px}. The
most general approach to performance forecasting involves the use of a
suitably defined utility function in the Bayesian framework, and it has
recently been presented in \cite{Trotta:2010ug}.

The purpose of this paper is to expand the FoM formalism to consider a new
dimension of the performance of a future dark energy probe, which has been
until today largely neglected -- namely, its robustness to potential
systematic errors. It is well known that systematic errors are going to be one
of the most challenging factors limiting the ultimate statistical performance
of precision measurements of $w(z)$. Yes there has been until now no formal
way to quantify how prone to potential systematic a future dark energy
measurement might be. This work takes a first step towards redressing this
issue, by introducing a so-called ``robustness'' FoM which complements the
statistical FoMs mainly considered so far in the literature.

This paper is organized as follows: in section~\ref{sec:fom} we introduce our
statistical and robustness FoMs, whose properties are illustrated in
section~\ref{sec:properties}. We then apply this formalism to future
supernovae type Ia and baryonic acoustic oscillation data in
section~\ref{darkenergy}. Our results and conclusions are presented in
section~\ref{sec:results}.

\section{Figures of Merit for Future Dark Energy Probes} \label{sec:fom}

\subsection{Gaussian linear model}

Suppose we have 2 different dark energy probes, whose likelihood function is assumed
to be Gaussian and is characterized by a Fisher matrix
$L_{i}$ ($i=1,2$), i.e. 
\begin{equation} \label{eq:likelihood_i}
\mathcal{L}_i (\Theta) \equiv p(d_i |\Theta)= \mathcal{L}_{0}^{i}
\exp\left(-\frac{1}{2}(\mu_{i}-\Theta)^{t}L_{i}(\mu_{i}-\Theta)\right).
\end{equation}
where $\Theta$ are the parameters one is interested in constraining and
$\mu_i$ is the location of the maximum likelihood value in parameter space.
In the absence of systematic errors, the maximum likelihood point, $\mu_{i}$,
is located at the true parameters value, which can be taken to be the
origin. We are here neglectic realization noise, i.e. we are working in the
Fisher matrix framework, in which $\mu_i$ is interpreted as the expectation
value of the maximum likelihood estimator averaged over many data
realizations. However, the presence of unmodelled systematic errors would
introduce a non--zero shift in $\mu_{i}$.  Below, we will show how the
systematic shifts $\mu_{i}$ can be estimated by propagating onto the parameter
space the shifts resulting from plausible unmodelled systematics in the
observables for dark energy.

The posterior distribution for the parameters, $p(\Theta|D)$, is obtained by
Bayes theorem as
\begin{equation}
p(\Theta|D)=\frac{p(\Theta)p(D|\Theta)}{p(D)},
\end{equation}
where $p(\Theta)$ is the prior, $p(D)$ the Bayesian evidence and $D$ are the data being used. 
If we assume a Gaussian prior centered on the origin with Fisher matrix $\Sigma$, the posterior from each probe is also a Gaussian, with Fisher matrix 
\begin{equation}
F_i = L_i + \Sigma \quad (i=1,2)
\end{equation}
and posterior mean 
\begin{equation} \label{eq:postmean}
\overline{\mu}_{i} = F_{i}^{-1}(L_{i}\mu_{i}).
\end{equation}
If we combine the two probes, we obtain a Gaussian posterior with Fisher matrix
\begin{equation} \label{eq:post_Fisher}
F = L_1 + L_2 + \Sigma
\end{equation}
and mean  
\begin{equation}
\mu =F^{-1}\sum_{i=1}^2 L_{i}\mu_{i}.  \label{eq:post_mean}
\end{equation}
 Notice that the precision of the posterior (i.e., the Fisher matrix) does not
 depend on the degree of overlap of the likelihoods from the individual
 probes. This is a property of the Gaussian linear model. In the presence of
 systematics, a FoM based on the posterior Fisher matrix is thus insufficient
 to quantify the power of the experiment: a future probe subject to systematic
 bias would have the same statistical FoM as an unbiased experiment. This
 motivates us to extend our considerations to a second dimension, namely
 robustness. 

For future reference, it is also useful to write down the general expression
for the Bayesian evidence. For a Normal prior
$p(\Theta)\sim{\mathcal{N}}(\theta_{\pi},\Sigma)$ and a
likelihood 
\begin{equation}
  \mathcal{L}(\Theta)=\mathcal{L}_{0}
  \exp\left(-\frac{1}{2}(\theta_{0}-\Theta)^{t}L(\theta_{0}-\Theta)\right),\end{equation}
the evidence for data $d$ is given by \begin{equation}
\begin{split}p(d) & \equiv\int{\rm {d}}\Theta p(d|\Theta)p(\Theta)=
  \mathcal{L}_{0}\frac{|\Sigma|^{1/2}}{|F|^{1/2}}\\
 & \exp\left[-\frac{1}{2}\left(\theta_{0}^{t}L\theta_{0}+
    \theta_{\pi}^{t}\Sigma\theta_{\pi}-\overline{\theta}^{t}F\overline{\theta}\right)\right],\end{split}
\label{eq:evidence}
\end{equation}
where $F = L + \Sigma$ and
$\overline{\theta} = F^{-1}(L\theta_0 + \Sigma \theta_\pi)$.

\subsection{The Statistical Figure of Merit}

It has become customary to describe the statistical power of a future dark
energy probe by the inverse area of its covariance matrix. This measure of
statistical performance  -- widely known as the DETF
FoM~\citep{Albrecht:2006um,HutererTurner} -- is usually defined (up to multiplicative
constants) as 
\begin{equation} \label{eq:DETF}
|L_i|^{1/2}.
\end{equation}  Here, we suggest to adopt a more statistically motivated measure of
the information gain, namely the Kullback-Leibler divergence (KL) between the
posterior and the prior, representing the information gain
obtained when upgrading the prior to the posterior via Bayes theorem:
\begin{equation} 
D_{KL} \equiv\int
  p(\Theta|D)\ln\frac{p(\Theta|D)}{p(\Theta)}d\Theta.
\label{eq:def_KL}
\end{equation}
The KL divergence measures the relative entropy between the two distributions: it is a dimensionless quantity which expressed the information gain obtained via the likelihood. 
For the Gaussian likelihood and prior introduced above, the information gain (w.r.t. the prior $\Sigma$) from the combination of both probes is given by
\begin{equation} \label{eq:DKL}
  D_{KL}=\frac{1}{2}\left(\ln|F|-\ln|\Sigma|-{\rm {tr}[1-\Sigma
      F^{-1}]}\right).
\end{equation} 
 
Below, we shall be interested in assessing the statistical performance of
future dark energy probes, in a context where probe 1 is taken to represent
present-day constraints on dark energy parameters, while probe 2 is a future
dark energy mission. We normalize the KL divergence for the combination of
probe 1 and probe 2, given by Eq.~\eqref{eq:DKL}, w.r.t. the case where probe
2 is assumed to be a hypothetical experiment that would yield the same dark
energy constraints as the existing ones (probe 1). This is not meant to
represent a realistic dark energy probe, but merely to give a benchmark
scenario for the normalization of the information gain.  This choice of
normalization has the added advantage of cancelling out most of the prior
dependence in Eq.~\eqref{eq:DKL}. After exponentiating the normalized KL
divergence, we therefore suggest to adopt as a statistical FoM the
dimensionless quantity
\begin{equation} \label{eq:S}
\begin{aligned}
S \equiv& \frac{|L_1+L_2+\Sigma|^{1/2}}{|2L_1 +\Sigma|^{1/2}} \\ 
 &\times \exp \left( \frac{1}{2}\rm {tr}[\Sigma((L_1+L_2+\Sigma)^{-1}-(2L_1+\Sigma)^{-1})] \right).
\end{aligned}
\end{equation}
%
%in the limit that $\Sigma << F_1, F_2$, then (I think) we can write:
%%
%\begin{align}
%S \equiv& \frac{|F_1+F_2|^{1/2}}{|2F_1|^{1/2}} \\  
% &\times \exp \left( \frac{1}{2}\rm {tr}[\Sigma((F_1+F_2)^{-1}-(2F_1+\Sigma)^{-1})] \right).
%\end{align}

%\dragan{You meant
%\begin{eqnarray*}  \label{eq:statFoM}
%S &\equiv& \frac{|F_2|^{1/2}}{|F_1|^{1/2}} 
%\exp \left(-\frac{1}{2}\rm {tr}[1-\Sigma F_2^{-1}]+ \frac{1}{2}\rm
%     {tr}[1-\Sigma F_1^{-1}] \right)\\[0.2cm]
%& = & \frac{|F_2|^{1/2}}{|F_1|^{1/2}} 
%\exp \left(\frac{1}{2}\rm {tr}[\Sigma (F_2^{-1}-F_1^{-1})] \right)\\[0.2cm]
%& = & \frac{|F_2|^{1/2}}{|F_1|^{1/2}}\,
%\frac{\left |\exp \left(\Sigma F_2^{-1}\right )\right |^{1/2}}
%{\left |\exp \left(\Sigma F_1^{-1}\right )\right |^{1/2}}
%\end{eqnarray*} 
%where in the last line I used the identity $\exp({\rm tr}(A)) = \det\exp(A)$.
%}

%\begin{align}
%S \equiv& \frac{|L_1+L_2+\Sigma|^{1/2}}{|2L_1 +\Sigma|^{1/2}} \\ 
% &\times \exp \left( \frac{1}{2}\rm {tr}[\Sigma((L_1+L_2+\Sigma)^{-1}-(2L_1+\Sigma)^{-1})] \right).
%\end{align}
%%
%in the limit that $\Sigma << F_1, F_2$, then (I think) we can write:
%%
%\begin{align}
%S \equiv& \frac{|F_1+F_2|^{1/2}}{|2F_1|^{1/2}} \\  
% &\times \exp \left( \frac{1}{2}\rm {tr}[\Sigma((F_1+F_2)^{-1}-(2F_1+\Sigma)^{-1})] \right).
%\end{align}
%}
%\dragan{I don't follow Marisa's logic.. I think we want to do (probe 2)/(probe
%  1) here, rather than (probe 1 + probe 2)/(2* probe
%  1). But this all depends on how the KL will be used later in the paper, which
%I haven't processed yet...}

% --x-X- end of comments on Eq.10 -X-x--
\subsection{Robustness of Dark Energy Probes}

In order to quantify the robustness to potential systematics of a combination
of probes, we wish to derive a measure of the degree of consistency between
them. The gist of our new robustness FoM is that our confidence in the
robustness of a new dark energy probe is increased if it returns constraints
which overlap significantly with previously existing probes. If on the
contrary the new probe has a small degree of consistency with previous
experiments, this might point to either a failure of the underlying
theoretical model or to the presence of unmodelled systematics in the new
probe (or both). In the following, we focus on the latter hypothesis.

The idea is to perform a Bayesian model comparison between two hypotheses,
namely $\mathcal{H}_{0}$, stating that the data $D$ are all compatible with
each other and the model, versus $\mathcal{H}_{1}$, purporting that the
observables are incompatible and hence tend to pull the constraints in
different regions of parameter space. The Bayes factor between the two
hypotheses, giving the relative probabilities (odds) between $\mathcal{H}_{0}$
and $\mathcal{H}_{1}$ is given by 
\begin{equation}
  R=\frac{p(D|\mathcal{H}_{0})}{\prod_{i=1}^2p(d_{i}|\mathcal{H}_{1})},\label{eq:rtest}\end{equation}
where the Bayesian evidence for a given hypothesis $\mathcal{H}$
is \begin{equation} p(d|\mathcal{H})=\int{\rm {d}}\Theta
  p(d|\Theta,\mathcal{H})p(\Theta|\mathcal{H}).
\end{equation} If $R \gg 1$,
this is evidence in favour of the hypothesis $\mathcal{H}_{0}$ that the data
are compatible. If instead $R \ll 1$ the alternative hypothesis
$\mathcal{H}_{1}$ is preferred (namely, that the data are incompatible). Examples of the application of the statistics $R$ introduced above can be found in~\cite{Hobson:2002zf,Feroz:2008wr} -- see the Appendix of~\cite{Feroz:2009dv} for a toy model illustration. For a review of Bayesian methods in cosmology, and in particular of model selection techniques, see \cite{Trotta:2008qt}.

We can restrict our considerations to just two probes, hence $D=\{d_{1},d_{2}\}$. Then the criterium of
Eq.~\eqref{eq:rtest} can be written as (omitting for brevity the explicit
conditioning on hypotheses)
\begin{equation}\label{eq:simplify_R}
R=\frac{p(d_{1},d_{2})}{p(d_{1})p(d_{2})}=\frac{p(d_{2}|d_{1})p(d_{1})}{p(d_{1})p(d_{2})}=
\frac{p(d_{2}|d_{1})}{p(d_{2})}.
\end{equation}
 The conditional evidence for $d_{2}$ given dataset
$d_{1}$ can be calculated as 
\begin{equation}
p(d_{2}|d_{1})=\int p(d_{2}|\Theta)p(\Theta|d_{1})d\Theta,
\label{eq:conditional_evidence}
\end{equation}
 where the first term is the likelihood for the second probe and the second
 term is the posterior from the first probe. By using the likelihood \eqref{eq:likelihood_i}, 
 and making use of Eq.~\eqref{eq:evidence} we obtain: 
\begin{equation}
\begin{split}p(d_{2}|d_{1}) & =\mathcal{L}_{0}^{(2)}\frac{|F_{1}|^{1/2}}{|F|^{1/2}}\\
 & \exp\left[-\frac{1}{2}\left(\mu_{2}^{t}L_{2}\mu_{2}+
    \overline{\mu}_{1}^{t}F_{1}\overline{\mu}_{1}-\mu^{t}F\mu\right)\right],\end{split}
\label{eq:pd2d1}
\end{equation}
 where $\mu$ is given by Eq.~\eqref{eq:post_mean}, $F$ by Eq.~\eqref{eq:post_Fisher} and $\overline{\mu}_{1}$ by Eq.~\eqref{eq:postmean}.
Using again Eq.~\eqref{eq:evidence} we obtain for the denominator
in Eq.~\eqref{eq:simplify_R} \begin{equation}
p(d_{2})=\mathcal{L}_{0}^{(2)}\frac{|\Sigma|^{1/2}}{|F_{2}|^{1/2}}
\exp\left[-\frac{1}{2}\left(\mu_{2}^{t}L_{2}\mu_{2}-\overline{\mu}_{2}^{t}F_{2}\overline{\mu}_{2}\right)\right],
\label{eq:pd2}
\end{equation}
 so that we obtain \begin{equation}
\begin{split}R & =\frac{|F_{1}|^{1/2}|F_{2}|^{1/2}}{|F|^{1/2}{|\Sigma|^{1/2}}}\\
 & \exp\left[-\frac{1}{2}\left(\overline{\mu}_{1}^{t}F_{1}\overline{\mu}_{1}+
    \overline{\mu}_{2}^{t}F_{2}\overline{\mu}_{2}-\mu^{t}F\mu\right)\right].\end{split}
\end{equation}
 Therefore we can recast Eq.~\eqref{eq:rtest}  into 
\begin{equation}
\begin{split} \ln R & =\frac{1}{2}\mu^{t}F\mu-\frac{1}{2}\sum_{i=1}^{2}\overline{\mu}_{i}^{t}F_{i}\overline{\mu}_{i}\\
 & -\frac{1}{2}\ln\frac{|F|}{|\Sigma|}+\frac{1}{2}\sum_{i=1}^{2}\ln\frac{|F_{i}|}{|\Sigma|}.\end{split}
\label{eq:FoM_2data}
\end{equation}
We shall use below the robustness $R$ to define a new FoM. For now, let us notice that it is the
product of two terms: the terms involving determinants of the Fisher matrices
add up to an Occam's razor factor, which is always $>0$. The second part
(summing over quadrating forms involving the various Fisher matrices)
expresses the degree of overlap of the posteriors from the two probes. This term will
reduce $R$ if the posteriors from the two probes are significantly
displaced from the posterior obtained using the combined data set (a smoking gun for systematic bias).  The
generalization of Eq.~\eqref{eq:FoM_2data} to an arbitrary number of probes is
derived in the Appendix.

% (remembering that $\ln R <0$ is evidence for incompatibility).

%The Occam's razor term, on the other hand, always tends to favour $\ln
%{R}>0$. Notice however that this term has only a mild, logarithmic dependence
%on the determinants, while the quadratic forms are generally much
%stronger. The degree by which the alternative hypothesis (that $d_{1}$ and
%$d_{2}$ need to be accounted for by including 2 distinct sets of parameters)
%is favoured by this term is proportional to the inverse of the prior
%determinant, $1/|\Sigma|$, just as in the usual calculation of the evidence
%for nested models~\cite{Trotta:2005ar}.  This arises because under the
%alternative hypothesis one has to introduce a second set of parameters to
%describe the second set of data $d_{2}$, and this is penalized because of the
%extra complexity this entails.  The amount of penalty is proportional to the
%volume under the posterior divided by the volume under the prior (i.e.,
%$\ln|F_{i}|/|\Sigma|$ terms). This entails a fixed strength of Occam's razor
%penalty, which does not depend on the degree of overlap of the probes.

\subsection{The Robustness Figure of Merit}

We now specialize to the situation where probe 1 describes our current
knowledge about dark energy parameter, while probe 2 represents a proposed
future dark energy mission. Notice that probe 1 does not need to be a single
experiment (i.e., just SN Ia or just BAO), but it can be interpreted as being
the effective joint constraint from a combination of all available present-day
dark energy probes. Without loss of generality, we assume that the current
constraints are unbiased, i.e. we set $\mu_{1}=0$ in the following, and we
wish to evaluate the robustness of a future probe, as defined in
Eq.~\eqref{eq:FoM_2data}, which might be subject to systematic bias.

Let us assume for the moment being that we can estimate the bias $b$ in
parameter space which probe 2 might be subject to. A procedure to achieve this
will be presented below for the specific cases of SN Ia and BAO
observations. For now, we remain completely general, and assume that the
maximum likelihood estimate for the dark energy parameters from probe 2 is
displaced from their true value by a bias vector $b$, i.e.~$\mu_{2} = b$.
This, together with the assumption that probe 1 is unbiased (i.e.,
$\mu_{1}=0$) gives $\bar{\mu}_{2} =F_{2}^{-1}L_{2}b$ and the joint posterior
mean from both probes is
\begin{equation}
\mu = F^{-1}L_{2}b.
\end{equation}
 Then we can write for the robustness $R$, Eq.~\eqref{eq:FoM_2data}
 \begin{equation}
\begin{split} \ln R & =\frac{1}{2}(F^{-1}L_{2}b)^tF(F^{-1}L_{2}b)-
  \frac{1}{2}(bL_{2})^tF_{2}^{-1}(L_{2}b)\\
  & -\frac{1}{2}\ln\frac{|F|}{|\Sigma|}+
  \frac{1}{2}\sum_{i=1}^{2}\ln\frac{|F_{i}|}{|\Sigma|},\end{split}
\label{gensigma}
\end{equation}
which can be rewritten as
 \begin{equation} \label{eq:lnR}
\begin{split} \ln R  & =-\frac{1}{2}(bL_{2})^t(F_{2}^{-1}-F^{-1})(L_{2}b)+\mathcal{R}_{0}\\
 & =-\frac{1}{2}b^tF^{*}b+\mathcal{R}_{0}\end{split}
\end{equation}
 where we have defined
 \begin{align} 
 F^{*} & \equiv L_{2}(F_{2}^{-1}-F^{-1})L_{2} \\
\mathcal{R}_{0} & \equiv -\frac{1}{2}\ln \frac{|F|}{|F_{1}|}\frac{|\Sigma|}{|F_{2}|}.
\end{align}
If the prior $\Sigma$ is negligible with respect to $L_{2}$ we have $F_{2}=L_{2}$
and $F^{*}=F_{2}-F_{2}F^{-1}F_{2}$. 

In order to normalize the value of $R$, we adopt the `repeated experiment' procedure we used for the normalization of the statistical FoM. This is
defined as the hypothetical case where the new experiment (probe 2) yields exactly the
same Fisher matrix as the existing probe 1, and is unbiased, i.e. $F_1=F_2$ and
$b=(0,0)$. For this identically repeated case the robustness of the two probes is given by 
\begin{equation}
R_* =\frac{|F_1|}{(|2L_1+\Sigma||\Sigma|)^{1/2}}. 
\label{eq:lnrstarfull}
\end{equation}

Normalizing $R$ from Eq.~\eqref{eq:lnR} to the above value means that
$R/R_*=1$ is the robustness that one would achieve by carrying out a
new dark energy measurement that would yield exactly the same
constraints as we currently have, and no bias. We therefore define the
quantity
\begin{equation} \label{eq:R_N}
R_N \equiv  \frac{R}{R_*}
\end{equation}
as our robustness FoM, which expresses the robustness of probe 2 under the assumption that it will be affected by a bias $b$. 

The robustness FoM above represents a ``worst case scenario'' (for a given
$b$) for probe 2, because we are assuming that it will be for sure
systematically biased. A more balanced approach is to average the robustness
along the direction defined by the systematic bias vector $b$. This gives an
``average robustness'', which accounts for different possible sizes in the
strength of the bias\footnote{Notice that as we average $R$ along $b$ we do
  not re-evaluate the Fisher matrix of the probe as a function of $b$, but we
  simply translate the Fisher matrix found at the fiducial point (i.e., the
  true parameters values).  The Fisher matrix typically depends only weakly on
  the fiducial model chosen, as long as we consider the models within the
  parameter confidence region.  If the bias vector is not much larger than the
  statistical errors we can therefore approximate the Fisher matrix at the
  biased parameters values with the one evaluated at the fiducial point.}.  In
order to perform the average, we rotate the coordinate axes so that the new
$x$-axis is aligned with the vector $b$ (assuming here a 2-dimensional
parameter space for simplicity):
\begin{equation}
\left( \begin{array}{c}1 \\ 0 \end{array}\right) =\Lambda b_{}
\end{equation}
where $\Lambda$ is a suitable rotation matrix, and $\Lambda^{t}=\Lambda^{-1}$.
Then the average robustness along the direction defined by $b$ is given by
\begin{equation} \label{eq:AvR}
\avR \equiv\int  W(x) \frac{R}{R_*}dx=\frac{e^{\mathcal{R}_{0}}}{R_*}\int W(x) e^{-\frac{1}{2}D_{11}x^2 } d x
\end{equation}
where
\begin{equation}
D \equiv \Lambda F^{*}\Lambda^{t}
\end{equation} 
and $W(x)$ is a suitable weighting function. A natural choice for $W$ is a
Gaussian with characteristic scale for the bias given by the length of the bias vector, $|b|$,
\begin{equation}
W(x)=\frac{1}{\sqrt{2\pi}|b|} e^{-\frac{1}{2}\frac{x^{2}}{|b|^{2}}},
\end{equation} 
so that Eq.~\eqref{eq:AvR} becomes
\begin{equation}
\begin{aligned} 
\label{eq:R_G}
\avR & = \frac{e^{\mathcal{R}_{0}}}{R_*\sqrt{2\pi}|b|}
\int e^{-\frac{1}{2}x^2 \left[ D_{11} + {|b|^{-2}}\right]}dx\\[0.2cm]
 & =   \frac{(|F_2| |2L_1 + \Sigma|)^{1/2}}{|F F_1|^{1/2}}||b|^{2}D_{11}+1|^{-1/2},
\end{aligned}
\end{equation}
 The Gaussian weight is centered at the unbiased parameter values, but it also has a tail that stretches above the characteristic scale of the bias, $|b|$, in order to account for a potentially much larger bias. We have checked that the use of other weight functions (e.g., a top-hat weight out to a maximum bias value given by the size of the bias vector) give a qualitatively similar result. We define the quantity given by Eq.~\eqref{eq:R_G} as the ``average robustness'' FoM. 
  
Finally, we can also combine the statistical and robustness FoMs to obtain an overall FoM expressing both the statistical power and the robustness to systematic bias of the probe as 
\begin{align}
T_{N} & \equiv R_N \times S \label{eq:T_N}, \\
\langle T \rangle & \equiv \avR \times S \label{eq:Tav}, 
\end{align}
where $S$ is given by Eq.~\eqref{eq:S}, while $R_N$ and $\avR$ by Eqs.~\eqref{eq:R_N} and \eqref{eq:AvR}, respectively.

\section{Properties of the Robustness FoM} \label{sec:properties}

Before applying the above formalism to future dark energy probes, we wish to
gain some further insight into the behaviour of our robustness FoM by
considering it in the context of a Gaussian toy model. We start with the normalized expression for the average robustness, Eq.~(\ref{eq:R_G})
and assume now that the confidence regions of the two probes are identical up to a roto-translation
(and therefore the determinants of $F_1,F_2$ are equal).
If moreover the prior is very weak we can approximate the posterior with the likelihood, hence 
\begin{equation}
\avR \approx 2 \frac{|F_{1}|}{|F|}||b|^{2}D_{11}+1|^{-1/2}.
\end{equation}
Let us further assume that probes 1 and 2 are aligned, i.e., they have
a degeneracy direction lying along the same straight line.  This means also that their Fisher matrices are simultaneously
diagonalizable (i.e. they commute) and that $F$ is also diagonalizable. Since the 
bias vector $b$ by definition connects the maximum likelihood points of the two probes, its direction is also aligned with one of the principal axis of the probes in this particular example. Then we can write
\begin{align}
D & =\Lambda(F_{2}-F_{2}F^{-1}F_{2})\Lambda^{-1}\\
 & =F_{2}^{D}-F_{2}^{D}(F_{1}^{D}+F_{2}^{D})^{-1}F_{2}^{D}\end{align}
where the superscript $D$ denotes the diagonalized version of a matrix.
The last step follows because for any matrix $A$ diagonalized by $\Lambda$
and any power $k$ one has\begin{equation}
\Lambda A^{k}\Lambda^{-1}=(A^{D})^{k}.
\end{equation}
Now let us denote the length of the $j$-th semiaxis of the $i$-th probe by $\sigma_{i,j}$
where (after diagonalization) the semiaxis $j=1(2)$ lies along the
abscissa (ordinate) . Then we have\begin{equation}
D_{11}=\sigma_{2,1}^{-2}\left(1-\frac{\sigma_{1,1}^{2}}{\sigma_{1,1}^{2}+\sigma_{2,1}^{2}}\right)\end{equation}
and therefore
\begin{align}
\avR & \approx \frac{2(\sigma_{2,1}\sigma_{2,2})}{(\sigma_{1,2}^{2}+
\sigma_{2,2}^{2}){}^{1/2}(|b|^{2}+\sigma_{1,1}^{2}+\sigma_{2,1}^{2})^{1/2}}.
\end{align}
This expression shows that the average robustness is invariant with respect to
rescaling of the axes: in fact, if the distances along the abscissa, $ \sigma_{1,1},\sigma_{2,1},|b|$,
are rescaled by an arbitrary factor, $\avR $ does not change; the same applies in the $y$-direction.

Since we assumed the ellipses to be congruent, we have two qualitatively
different cases: orthogonal ellipses ($\bot$), i.e.
$\sigma_{2,2}=\sigma_{1,1}$ and $\sigma_{2,1}=\sigma_{1,2}$; and parallel
ellipses ($\parallel$), i.e.  $\sigma_{1,1}=\sigma_{2,1}$ and
$\sigma_{1,2}=\sigma_{2,2}$.  In the orthogonal case we obtain
\begin{align}
\avR^{\bot} & =\frac{2r}{1+r^2}\left(1+\frac{|b|^2 r^2}{\sigma_{2,1}^2(1+r^2) }\right)^{-1/2}
\end{align}
where $r=\sigma_{2,1}/\sigma_{2,2}$ measures the elongatedness of the ellipses.
In the parallel case we obtain instead for any $r$
\begin{align}
\avR^{\parallel} & =\left(1+\frac{|b|^2}{2\sigma_{2,1}^2 }\right)^{-1/2}
\end{align}
From these expressions we can derive some general consequences. Because of our
choice of normalization, unbiased identical probes have unity robustness. In
general, if the bias length is small with respect to the statistical errors of
the second probe, then parallel probes are more robust than orthogonal
ones. If the second probe is very elongated (degenerated) along the bias
direction, i.e.  $r\gg 1$, then again parallel probes are more robust than
orthogonal ones. If instead the degeneracy of the second probe lies
orthogonally to the bias direction, $r\ll 1$, there are two cases: parallel
probes are more robust if the bias is smaller than the long axis
($|b|^2<\sigma_{2,2}$), but less robust in the opposite case.  Of course the
general case, with arbitrary bias direction and length and arbitrary sizes and
orientation of the probes cannot be reduced to such simple conclusions.

\begin{figure}
\begin{centering}
\includegraphics[width=0.49\linewidth]{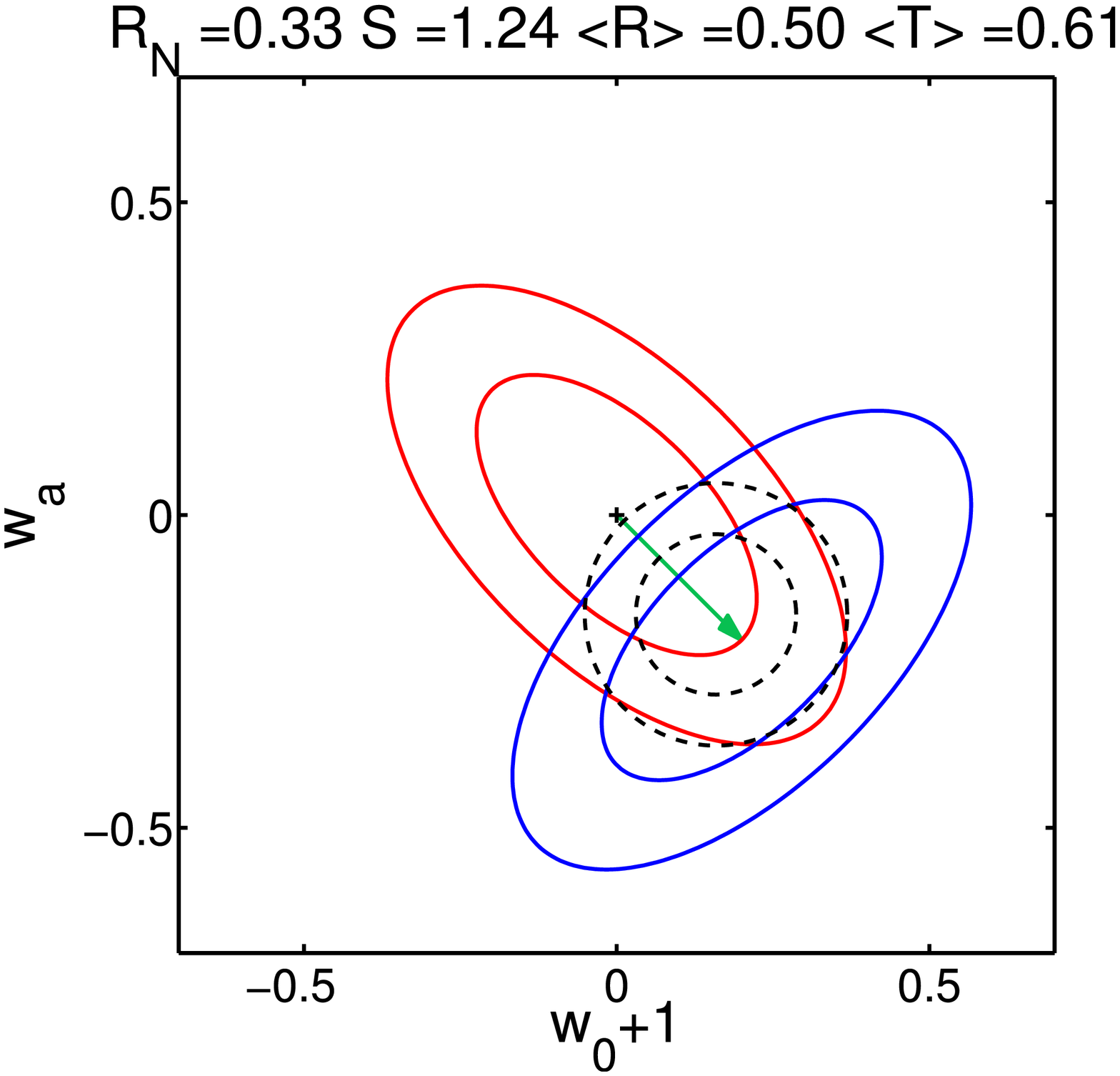}\hfill{}
\includegraphics[width=0.49\linewidth]{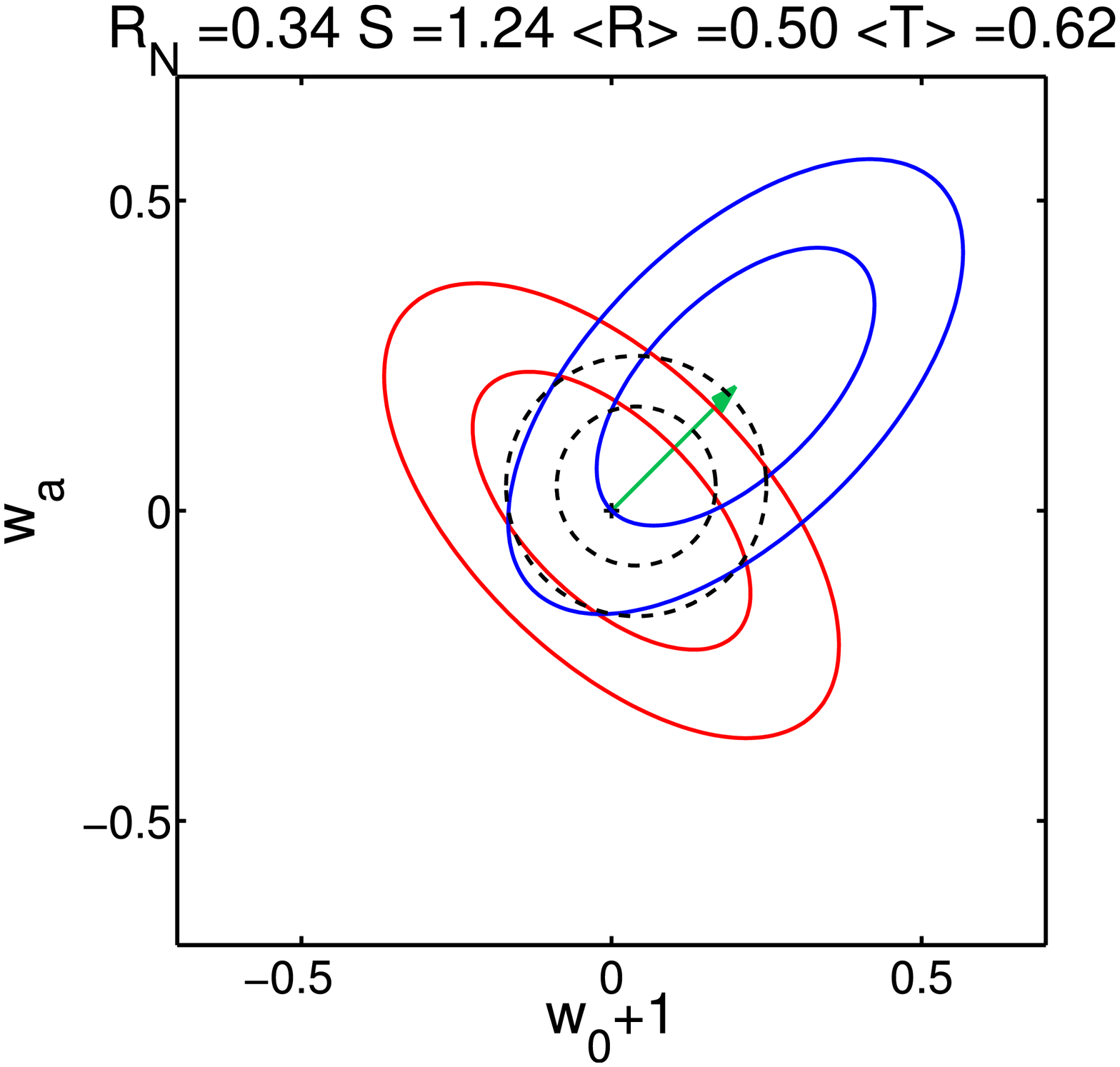}\vfill{}
\includegraphics[width=0.49\linewidth]{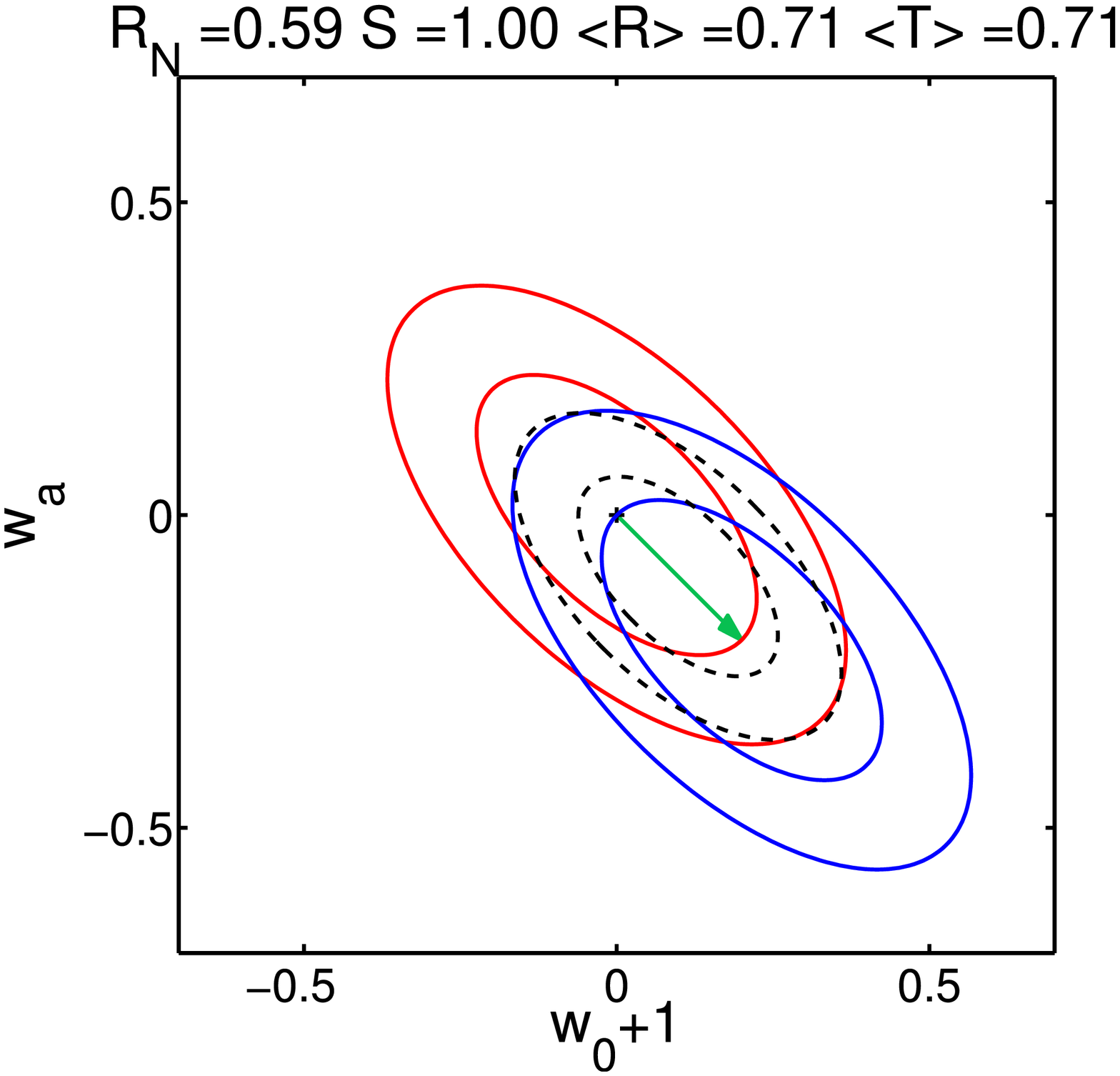}\hfill{}
\includegraphics[width=0.49\linewidth]{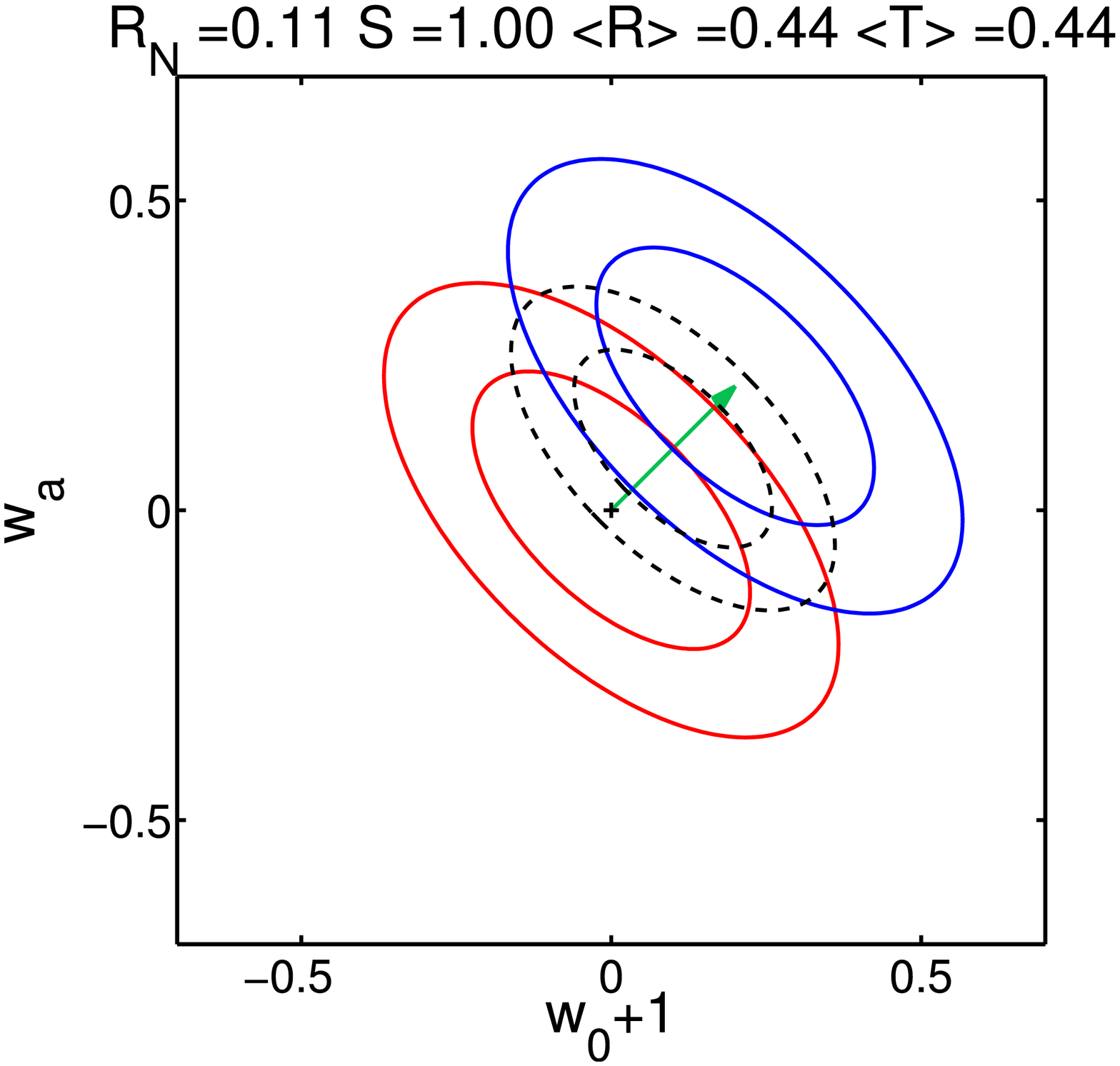}
\par\end{centering}
\caption{Illustration of statistical and robustness FoM for a future probe
  (blue ellipses, 68\% and 95\% C.L.) which is systematically biased w.r.t. the present-day
  constraints (red ellipses) in the direction given by the green bias
  vector. The black dotted ellipses represent the combined constraints. Notice that the statistical FoM (S) does not change in the presence of a systematic bias.}
\label{fig:toy1} 
\end{figure}

Armed with the above intuition, we now consider in Fig.~\ref{fig:toy1}, a
numerical illustration of 4 different cases for the relative orientation of
the two probes (orthogonal or parallel) and the direction of the bias vector
(along the short or long semiaxis). The two sets of iso-likelihood contours
enclose 68\% and 95\% confidence levels; as above, the second probe (blue
contours) has the same area as the first (i.e., $L_1 = L_2$), but its
degeneracy direction can be rotated, and its maximum likelihood value is
displaced from the true value by a systematic bias (of fixed length in all
cases), given by the green vector. The first probe (red contours) is assumed
to be unbiased.  The prior is $\Pi = \text{diag}(1,1)$ (i.e.\ a prior of 1.0
in each parameter with no correlations imposed).  For each case, we give the
corresponding statistical FoM, Eq.~\eqref{eq:S}, the robustness
FoMs,~\eqref{eq:R_N} and \eqref{eq:AvR}, and the total FoM (for the averaged
robustness), Eq.~\eqref{eq:Tav}.

The robustness FoM (with or without averaging) depends both on the
direction along which the bias is directed and on the relative
orientation of the degeneracy directions of the two probes. When the
bias is directed along the degeneracy direction of probe 1 and probe 2
is aligned along that direction (lower left panel), the robustness is
maximal. It decreases if the two probes are orthogonal to each other,
since this reduces the degree of overlap between them (upper
panels). Finally, robustness is smallest when the two probes are
aligned but the bias is direct orthogonally w.r.t the degeneracy
direction (lower right panel), as argued above. Looking ahead
  to the application of the robustness formalism to the dark energy
  equation of state parameters in the next section, we can anticipate
  here that the most relevant case is the one where the two probes are
  similarly oriented (bottom panels of Fig.~\ref{fig:toy1}). This is
  because different dark energy probes are typically degenerate in the
  equation of state parameters along quite similar
  directions. Therefore, their relative robustness can be expected to
  depend mainly on the orientation of the bias w.r.t.~the main
  degeneracy direction.

The statistical FoM is largest when the probes are
orthogonal to each other, as expected. Notice that the statistical FoM is
unaffected by the bias, and only depends on the relative alignment of the two
probes. For a given orientation and size of the bias vector, the total FoM
allows one to decide which configuration for probe 2 is to be preferred. For
the example of Fig.~\ref{fig:toy1}, if the bias vector points along the
degeneracy direction of probe 1 (left-hand side panels), one would prefer
probe 2 to be aligned with probe 1 ($\avT = 0.71$) as opposed to probe 2 being
orthogonal to probe 1 ($\avT = 0.61$). If instead the bias is orthogonal to
the degeneracy of probe 1 (right-hand side panels), then the best choice for
probe 2 is for it to be orthogonal to probe 1 ($\avT = 0.62$ compared to $\avT
= 0.44$).

% -----x-X- BEGIN: Toy model fig:  orthogonal bias -X-x-----
\begin{figure}
\begin{centering}
\includegraphics[angle=0,width=\linewidth]{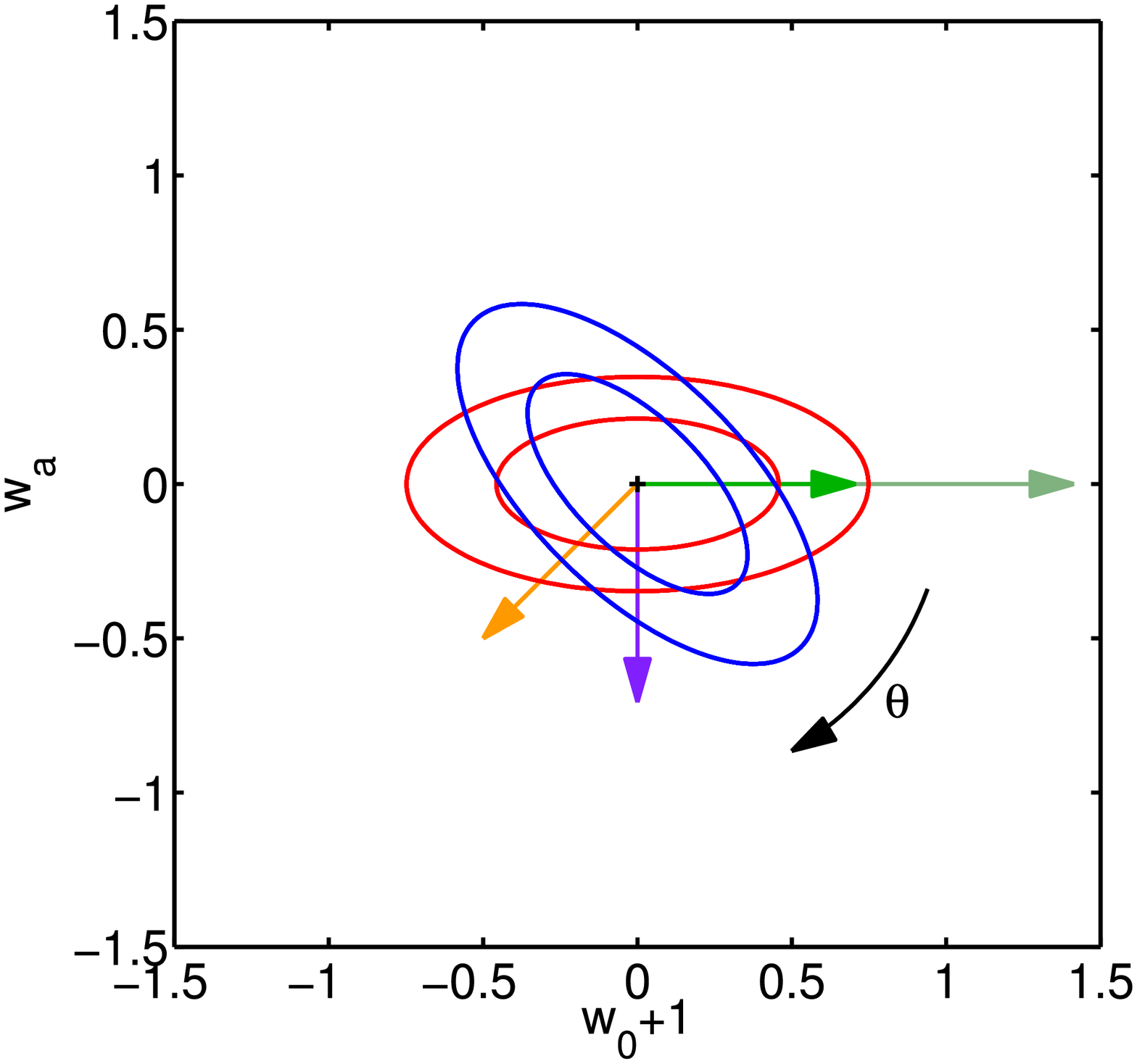} \\
\includegraphics[angle=0,width=\linewidth]{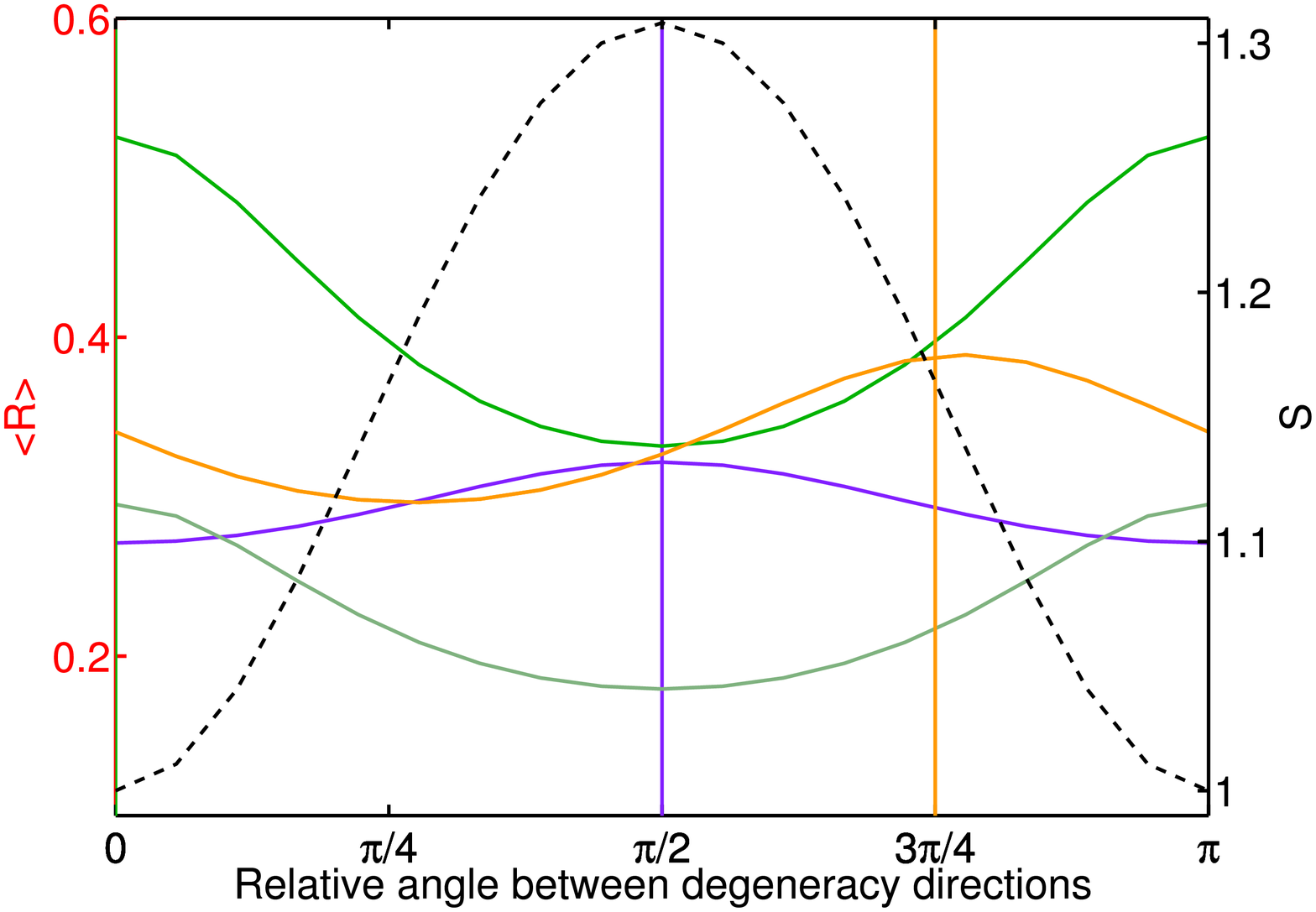} 
\par\end{centering}
\caption{Dependency of FoMs on the angle between the degeneracy direction of
  the two probes.  Upper panel: the red (blue) ellipses represent the 68\% and
  95\% likelihood contours of probe 1 (probe 2, which is potentially
  biased). The degeneracy direction of probe 2 is offset by an angle $\theta$
  w.r.t probe 1. The three vectors gives three possible directions for the
  bias. Lower panel: value of statistical FoM $S$ (black dashed line, right-hand
  axis), and average robustness FoM, $\avR$, (coloured solid lines, left-hand
  axis, colour and thickness matching the bias vectors in the upper panel), as a function of
  the relative angle $\theta$. Vertical coloured lines give the angle of each
  bias vector. \label{fig:toy_rot} }
\end{figure}

%-----x-X- END: Toy model fig:  orthogonal bias -X-x-----
We can also ask what is the optimal orientation of probe 2 with respect to
probe 1 if one wanted to maximise its robustness, given a bias direction.  In
Fig.~\ref{fig:toy_rot}, we plot both the statistical and the average
robustness FoMs as a function of the rotation angle between the principal
direction of the two probes. The average robustness is evaluated for 3
different directions of the bias (coloured vectors in the top panel). We
notice once more that the statistical FoM is maximised when the probes are
orthogonal.  However, the robustness FoM is maximised when the degeneracy
direction of probe 2 is close to being aligned with the direction of the bias
vector, as this maximizes the overlap with probe 1 even when probe 2 suffers
from a systematic error. Finally, increasing the length of the bias by a
factor of 2 (fainter green vector in the top panel) reduces the overall
average robustness.

In summary, the robustness of a future probe is a function of its
statistical properties (i.e., the direction along which its main
degeneracy is aligned, compared with the degeneracy direction of probe
1) as well as of the direction and size of the systematic bias. The
performance of a future probe should be assessed by considering
simultaneously its statistical power but also its robustness to
systematics. Optimizing a future dark energy experiment in terms of
its statistical errors alone would generically lead to an experiment
which is less robust, for a given overall level of plausible
systematics. Any optimization procedure should therefore involve the
complementary criteria of statistical strenght and robustness to
systematic bias.

%-----x-X- BEG FIGURE: ROTATION DEPENDENCY 10.12.2010 -X-x---------
%
%-----x-X- END FIGURE: ROTATION DEPENDENCY 10.12.2010 -X-x---------

We now turn to applying the above concept to the concrete scenario of two
classes of future dark energy missions, namely type Ia SN and BAO
measurements.

\section{Robustness of Future Dark Energy Probes}
\label{darkenergy}

We consider a
simple and widely used phenomenological description of an evolving dark energy
model, where the equation of state is $w(z) = w_0 + w_a z/(1+z)$, characterized by
the two free parameters $(w_0, w_a)$ \citep{CP2003,Linder_wa}. For probe 1 (representing
current constraints on $w_0, w_a$) we take a Gaussian approximation to the
joint likelihood resulting from the combination of Union 2 SNe Ia
data~\citep{AmanullahLidman2010}, SDSS BAO~\citep{PercivalReid2010}, WMAP7
measurements of the shift parameters~\citep{KomatsuSmith2010}, and SHOES
measurements of the Hubble constant~\citep{RiessMarci2009}. We further assume
a flat Universe. For the prior on the dark energy parameters, we take a
Gaussian centered at $(w_{0}+1,w_{a})=(0,0)$ with Fisher matrix $\Pi =
\text{diag}(1,1/100)$. In the following, when we look at the Fisher matrix we
mean the 2D marginalised Fisher matrix (i.e., marginalised down to the dark
energy parameter space). Although in the rest of this paper we focus exclusively on the robustness FoM for dark energy parameters, we note that our robustness formalism is equally applicable to any other cosmological parameter one is interested in. 

In order to evaluate robustness, we need to specify the bias vector $b$. There
are several plausible ways of doing this, and the outcome will depend on what
one thinks a possible systematic bias might be due to. In our case, in order
to illustrate our new FoM, we determine $b$ by assuming a possible systematic
bias in the probe's observables, and then projecting the resulting systematic
shift onto the dark energy parameter space of interest, as described in detail
below. We stress that this is by no means the only procedure by which one can
estimate $b$. Other assumptions about the origin of systematic errors will in
general lead to a different $b$, and therefore to a different value for the
robustness of the future probe.

\subsection{Future SN Ia Measurements}

We consider a survey dedicated to observing type Ia supernovae from space, with a redshift distribution like the one expected from SNAP, with 2000 SNe distributed as in \cite{KLMM_SNAP}, plus a
low-$z$ sample of 300 SNe distributed uniformly in the redshift range
$0.03<z<0.08$. The projected SNAP magnitude errors include both statistical
and systematic components, and are modelled as follows:
\begin{equation}
\sigma_{b}^{2}=\left[\frac{0.15^{2}}{N_{b}}+A_\text{syst}^{2}
\left(\frac{1+z_{b}}{1+z_\text{max}}\right)^{2}\right],
\label{eq:magerr}
\end{equation}
where $N_{b}$ is the number of SNe in each bin centered at $z_{b}$ and of
width $dz=0.1$. The second term on the right-hand side of
Eq.~(\ref{eq:magerr}) models a systematic floor that increases linearly with
$z$ up to a maximum of $A_\text{syst}$ mag per $dz=0.1$ bin at $z_{{\rm max}}
= 1.7$ \citep{LinHut_highz}. In order to evaluate the robustness of SNa data
for different levels of systematics, we will consider values of $A_\text{syst}
= 0.01, 0.02, 0.05$.

\begin{figure}
\begin{centering}
\includegraphics[width=0.8\linewidth,angle=-90]{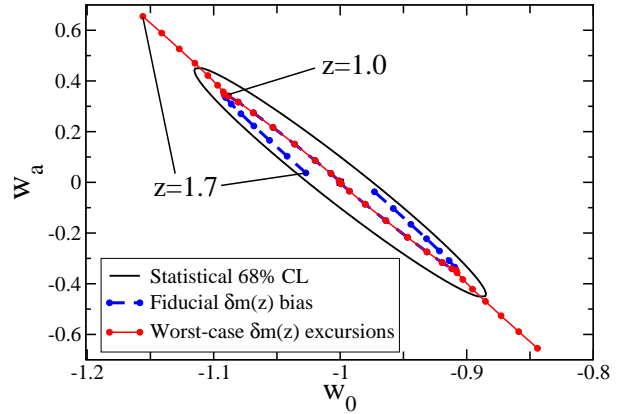} 
\par\end{centering}

\caption{Systematic bias in the $w_{0}$-$w_{a}$ plane for future SNIa
  data. The square denotes our fiducial value and the ellipse gives
  the 68\% CL statistical contour from future SNIa data. The blue
  curve shows the systematic bias given by Eq.~(\ref{eq:dm_bias}),
  with points showing cumulative contributions from each of the 16
  redshift bins -- that is, cumulative value of the sum in
  Eq.~(\ref{eq:dp_bias}).For clarity, we explicitly label
    bias contributions accumulated by redshifts $z=1$ and
    $z=1.7$. The red segments denote the worst-case
  bias, where the sign of $\delta m(z)$ at each redshift bin conspires
  to shift the $(w_{0},w_{a})$ value away from the true value
  (``Maximum excursion bias''); see Eq.~(\ref{eq:dp_worst_bias}). For
  clarity, we have also plotted the biases with the opposite sign
  relative to the fiducial model parameter values.  }
\centering{}\label{fig:w0wa_bias}
\end{figure}

We assume a flat universe with four parameters relevant for this analysis,
matter density relative to critical $\Omega_{M}$, equation of state
today $w_{0}$, its variation with scale factor $w_{a}$, and a nuisance
offset in the Hubble diagram $\mathcal{M}$. Marginalizing over $\Omega_{M}$
and $\mathcal{M}$ and assuming $A_\text{syst} = 0.02$, we find that our fiducial survey produces statistical
errors of $\sigma_{w_{0}}=0.075$ and $\sigma_{w_{a}}=0.30$, corresponding
to the black 68\% C.L.\ ellipse in Fig.~\ref{fig:w0wa_bias}.

The bias in the dark energy parameters, $b$, reconstructed from SN measurements induced by an arbitrary bias in the observed
magnitudes $\delta m(z)$ can be derived from the Fisher
matrix for SNe (e.g.\ \cite{Knox:1998fp,HutererTurner}), and is given by
\begin{eqnarray}
b_{i} & = & 
\sum_{j}(F^{-1})_{ij}\sum_{\alpha}\frac{dm(z_{\alpha})}{d\mu_{j}}\,
 \frac{1}{\sigma_{\alpha}^{2}}\,\delta m_{{\rm sys}}(z_{\alpha})\\[0.2cm]
 & \equiv & \sum_{\alpha}c_{\alpha}^{(i)}\,\delta m_{{\rm sys}}(z_{\alpha})
\label{eq:dp_bias}
\end{eqnarray}
where $\mu_{i}$ are the cosmological parameters,
$c_{\alpha}^{(i)}\equiv\sum_{j}(F^{-1})_{ij}(dm(z_{\alpha})/d\mu_{j})/\sigma_{\alpha}^{2}$
and where $\alpha$ runs over redshift bins. We adopt a systematic bias of the
same form as the ``floor'' that was previously included in the total
\textit{statistical} error:
\begin{equation} 
\delta m_{{\rm sys}}(z_{\alpha})=A_\text{syst}\left(\frac{1+z_{\alpha}}{1 + z_\text{max}}\right)
\label{eq:dm_bias}
\end{equation}
Bias of this magnitude leads to the bias on cosmological parameters which can
be calculated using Eqs.~(\ref{eq:dp_bias}) and (\ref{eq:dm_bias}), and is
shown as the blue curve in Fig.~\ref{fig:w0wa_bias}. Each point on the curve
shows cumulative contributions to the excursion in $w_{0}$ and $w_{a}$ around
the fiducial model (with $(w_0, w_a) = (-1,0)$) for each of the 16 redshift
bins we consider, $z\in[0.1,0.2],\ldots,z\in[1.6,1.7]$. In other words, points
on the blue curve show cumulative contribution of the sum in Eq.~(\ref{eq:dp_bias}).

But this form of the bias assumes that the excursions in $\delta m(z_\alpha)$ are of
the same sign (taken to be positive), and equal to the maximum allowed value
in Eq.~(\ref{eq:dm_bias}). The worst-case bias to the dark energy parameters
is obtained if $\delta m(z_\alpha)$ changes sign in each redshift bin just so as to
maximize the excursion in $w_{0}$ and $w_{a}$. Such a worst-case bias can be
straightforwardly calculated \citep{Huterer:2004tr}
\begin{equation}
b_i^{{\rm worst}}=\sum_{\alpha}|c_{\alpha}^{(i)}|\,\delta m_{{\rm
    sys}}(z_{\alpha})\label{eq:dp_worst_bias}
\end{equation}
for a single dark energy parameter $\mu_{i}$, where $\delta m_{{\rm
    sys}}(z_{\alpha})>0$ was taken by default. In other words, the systematic
error takes a plus or minus sign equal to that of the coefficient
$c_{\alpha}^{(i)}$ in each redshift bin
\footnote{For multiple parameters, there is ambiguity to define the worst-case
  error, since a sign of $\delta m_{{\rm sys}}(z_{\alpha})$ that makes
  excursion in $w_{0}$ positive may actually make the $w_{a}$ excursion
  negative or vice versa. We make a choice that the excursion in $w_{0}$ is
  positive in a given redshift bin, which determines the sign of $\delta
  m_{{\rm sys}}(z_{\alpha})$; then the excursion in $w_{a}$ in that bin is
  simply $c_{\alpha}^{(w_{a})}\,\delta m_{{\rm sys}}(z_{\alpha})$.}. Such a
  worst-case excursion in the $(w_{0},w_{a})$ plane is shown as the red curve
  with points in Fig.~\ref{fig:w0wa_bias}. We call this scenario the ``maximum
  excursion bias'' (MEB), and use it as an estimate for the bias vector $b$ in
  the computation of our robustness FoM.

\subsection{Future Baryonic Acoustic Oscillations Measurements}

The second class of future probe we consider consists of a full-sky
spectroscopic redshift survey modelling a future spacee mission with
specifications close to WFIRST or Euclid (or a Stage-IV mission in the
language of the DETF).  The probe is fully specified by choosing a
number of redshift bins and giving the expected number densities of
galaxies per bin and the sky coverage, assumed here to be 20,000
square degrees. Table \ref{tab:n_z} gives the redshift binning and the galaxy number densities,
taken from the data published by the Euclid collaboration
\citep{2009arXiv0912.0914L}. We assume however that only half of these
galaxies can be effectively employed (efficiency $\epsilon=0.5$), 
corresponding to one half of values in the Table.
\begin{table}
\begin{tabular}{c c c c}
\hline 
$z$  & $n(z)\times10^{-3}$  
\tabularnewline
\hline
\hline 
0.5-0.7   & $3.56$  \tabularnewline
0.7-0.9    & $2.42$  \tabularnewline
0.9-1.1    & $1.81$  \tabularnewline
1.1-1.3    & $1.44$   \tabularnewline
1.3-1.5   & $0.99$   \tabularnewline
1.5-1.7    & $0.55$  \tabularnewline
1.7-1.9    & $0.29$  \tabularnewline
1.9-2.1    & $0.15$  \tabularnewline
\hline
\end{tabular}

\caption{\label{tab:n_z}Expected galaxy number densities per redshift bin in
  units of ($h/$Mpc)$^{3}$ for the Euclid survey. }

\end{table}

In order to forecast the statistical errors on dark energy parameters,
we adopt the Fisher matrix method
of~\cite{2003ApJ...598..720S,2007ApJ...665...14S}, also employed in
\cite{2005MNRAS.357..429A}. Here we give a
short summary of the method and refer to these papers for the
implementation details.  In the limit where the survey volume $ \Vsur$
is much larger than the scale of any features in $P_{\rm obs}(k)$, it
has been shown that the redshift survey Fisher matrix in a redshift
bin $ \Delta z$  can be approximated as \citep{Tegmark97}
\begin{eqnarray}
F_{ij}
&=&\int_{-1}^{1} \int_{k_{\rm min}}^{\kmax}\frac{\partial \ln
  P_{\rm obs}(k,\mu)}{\partial \mu_i} \frac{\partial \ln P_{\rm obs}(k,\mu)}{\partial \mu_j}\nonumber\\
&\times & 
\Veff(k,\mu) \frac{2\pi k^2 dk d\mu}{2(2\pi)^3}
\label{Fisher}                 
\end{eqnarray}
Here, $k,\mu$ are the wavevector moduls and direction cosine with respect to the line of sight, 
respectively, and
 the derivatives are evaluated on the parameters $\mu_i$
of the fiducial model. The upper cut-off $\kmax$ is chosen so as to
avoid the non-linear regime, while the large-scale cut-off $\kmin$ is set to $0.001 h$/Mpc but
its precise value has a very weak impact.
 $\Veff$ is the effective volume of the survey:
\begin{eqnarray}
\Veff(k,\mu) =
\left [ \frac{{n_{\rm g}}P_{\rm g}(k,\mu)}{{n_{\rm g}}P_{\rm g}(k,\mu)+1} \right ]^2 \Vsur,
\label{V_eff}                                                                                        
\end{eqnarray}
where $\Vsur$ is the 20,000 square degrees survey volume  contained in a given redshift bin. The galaxy comoving number density
$n_{\rm g}(z)$ is assumed to be spatially constant within a redshift
bin, while $P_g$ is the galaxy spectrum defined below.  
The total Fisher matrix is obtained by summing over all the redshift bins of Table \ref{tab:n_z}.
The matter power
spectrum in any given cosmology can be written in terms of the
spectrum in the fiducial (or ``reference", subscript ``ref") cosmology
as
\begin{equation}
P_{\rm obs}(k_{{\rm ref}\perp},k_{{\rm ref}\parallel},z)
=\frac {\DA _{\rm ref} ^2 \hz}{\DA ^2 \hz _{\rm ref}} P_{\rm g}(k_{{\rm ref}\perp},k_{{\rm ref}\parallel},z)
+P_{\rm shot}\,,
\label{eq:Pobs}
\end{equation}
where
\begin{equation}
P_{\rm g}(k_{{\rm ref}\perp},k_{{\rm
    ref}\parallel},z)=b(z)^2\left[1+\beta(z) 
\frac{k_{{\rm ref}\parallel}^2}{k_{{\rm ref}\perp}^2+k_{{\rm ref}\parallel}^2}\right]^2
P_{{\rm matter}}(k,z)\,.
\label{eq:Pg}
\end{equation}
In Eq.~(\ref{eq:Pobs}), $H(z)$ and $D(z)$ are the Hubble parameter and
the angular diameter distance, respectively, and the prefactor
encapsulates the geometrical distortions due to the Alcock-Paczynski
effect \citep{2003ApJ...598..720S,2007ApJ...665...14S}.  $k_\perp$ and
$k_\parallel$ are the wave-numbers across and along the line of sight
in the given cosmology, and they are related to the wave-numbers
calculated assuming the reference cosmology by $k_{{\rm ref}\perp} =
k_\perp D(z)/D(z)_{\rm ref}$ and $k_{{\rm ref}\parallel} = k_\parallel
H(z)_{\rm ref}/H(z)$.  $P_{\rm shot}$ is the unknown white shot noise that
remains even after the conventional shot noise of inverse number
density has been subtracted
\citep{2003ApJ...598..720S,2007ApJ...665...14S}.  In
Eq.~(\ref{eq:Pg}), $b(z)$ is the \emph{linear bias} factor between
galaxy and matter density distributions,
$f_g(z)$ is the linear growth rate, $\beta(z)=f_g(z)/b(z)$ is the
linear redshift-space distortion parameter  \citep{Kaiser1987} and $P_{{\rm matter}}$ is the
linear matter power spectrum. The fiducial values for the bias and the growth factor are
  $b(z)=1$ and $f_g=\Omega_M^{0.545}$, respectively. In \cite{diPorto2010} it has been shown that the
precise fiducial value of $b(z)$ does not have a large impact on the results.

This method employs all the information
contained in the power spectrum, including the redshift distortion,
not just the position of the baryonic wiggles.  As above, we choose a
flat fiducial cosmology with $\Omega_{M}=0.24$, $h=0.7$,
$\Omega_{DE}=0.737$, $\Omega_{K}=0$, $\Omega_{b}h^{2}=0.0223$,
$n_{s}=0.96,$ $w_{0}=-1,w_{a}=0$.  Unlike in the SN
  case, we do not impose an explicit systematic floor in the forecasted BAO
  errors; the finite sky coverage and density of galaxies provide an
  effective floor for any given BAO survey.
As mentioned above, beside the cosmological
parameters, for each redshift bin we also include as free parameters
to be differentiated (and then marginalized) in the Fisher matrix a matter-galaxy bias factor
and an additive shot noise term in the power spectrum (for details see
\cite{2005MNRAS.357..429A}). These terms act as additional effective systematic floors.

The systematic effect we assume for the redshift survey is a
fractional error in estimating the value of the Hubble function
$H(z_i)$ of magnitude $A_\text{syst} = 0.001, 0.002, 0.005$ in each
bin $i$.  Such a bias in $H(z)$ propagates to a bias in the angular
diameter distance $D(z)$, as well, if the standard flat-space
Friedman-Robertson-Walker relation
\begin{equation}
D(z)=(1+z)^{-1}\int_0^z \frac{dz'}{H(z')}
\end{equation}
holds true, which we assume here. The angular diameter distance bias
is then related to the Hubble function bias by
\begin{equation} \label{eq:syst_shift_relation}
\delta (\ln D)=-\delta(\ln H)\frac{H(z)}{(1+z)D(z)}\int_0^z \frac{dz'}{H^2(z')}.
\end{equation}
where we have used the assumption that  the bias in $\ln H$ is redshift-independent.
This simple choice for modelling systematic errors in BAO is meant to
approximately capture a possible systematic shift in the baryon peak
position due to e.g. the presence of isocurvature modes~\citep{Zunckel:2010mm}
or non-linear effects, of the kind described e.g. in~\cite{seo2008}. A
more realistic choice of systematic errors is difficult to model accurately
(as, for example, a bias in $H(z)$ and/or $D(z)$ also modifies in general the
whole spectrum behavior and the redshift distortions), and it is left for
future work. Our present choice is meant as a simple illustration of the
method and a first step towards evaluating the robustness FoM.

 If instead of the true matter
power spectrum, $P(k)$, we measure a spectrum that contains a
systematic error $\delta s_{\alpha} = \delta (\ln H_{\alpha})$ or $\delta s_{\alpha} = \delta
(\ln D_{\alpha})$ in the value of $H(z_{\alpha})$ and $D(z_{\alpha})$ (where the systematic shifts
are related by Eq.~\eqref{eq:syst_shift_relation}), the maximum likelihood
estimate for the $i$-th parameter will be shifted w.r.t. its true value by a bias given
by (see e.g. \cite{2007MNRAS.374.1377T})
\begin{eqnarray}
\delta \mu_i & = & F_{i j}^{-1}\left [\frac{1}{8\pi^{2}}\int d\mu k^{2}dk
  \frac{\partial\ln P}{\partial\mu_{j}}\frac{\partial\ln P}{\partial
 s_{\alpha}}\right ]\delta s_{\alpha} \nonumber \\
 &\equiv & c^{(i)}_{\alpha} \delta s_{\alpha}
\label{eq:mubias-1}
\end{eqnarray}
(sum over repeated indexes).  Analogously to the previous
subsection we have defined
\begin{equation}
c^{(i)}_{\alpha}\equiv
  F_{i j}^{-1}\left [\frac{1}{8\pi^{2}}\int d\mu k^{2}dk
  \frac{\partial\ln P}{\partial\mu_{j}}\frac{\partial\ln P}{\partial
    s_{\alpha}}\right ]\,.
\end{equation}
In this particular case, however, the $i$-th parameters coincide with $\delta s_{i} = \delta (\ln H_i), \delta
(\ln D_i)$ and therefore the matrix $ c^{(i)}_{\alpha}$ is the identity matrix.
We can then directly project the systematic bias onto the dark energy parameters $(w_0, w_a)$, obtaining a bias vector $b$ of the form
\begin{equation}
\begin{split}
b_l & = 
\sum_\beta \left(\frac{\partial w_{l}}{\partial \ln H(z_\beta)} \right. - \\ 
& \frac{H(z_\beta)}{(1+z_\beta)D(z_\beta)}
\left. \int_0^{z_\beta} \frac{dz'}{H^2}\frac{\partial w_{l}}{\partial \ln D(z_\beta)}\right)\,\delta \ln H(z_\beta)
\end{split}
\end{equation}
 where $\delta \ln H(z_{\beta}) = 0.001, 0.002, 0.005$,
the subscript $\beta$ runs over the redshift bins, and $l=0,a$. We have chosen to
consider systematic shifts in the range of 0.1\% to 0.5\% to reflect ballpark
estimates of what BAO systematic errors due e.g. to residual non-linear
corrections might be. We stress once more that this is a simplified treatment
used here mainly for illustration purposes of our method.

We evaluate $b_{l}$ for each redshift bin and then estimate the maximum bias
by following the same method discussed in the previous subsection. Here it
happens that the contributions to $b_{l}$ are always positive and therefore
automatically select the worst case scenario, i.e., the maximum excursion
bias. The resulting maximum excursion bias for different levels of systematics
is shown as the green vectors in the bottom panel of Fig.
\ref{fig:BAO_SN_ave}, together with the statistical errors from our BAO probe
(blue ellipses) and current constraints (red ellipses), plotted for
comparison.

\section{Results}
\label{sec:results}

Our results for the statistical and robustness FoMs are summarized in
Table~\ref{tab:ave_results}, where we give the values of our robustness,
statistical and total FoM. We also show the value of the DETF FoM (normalized
to the value obtained from the current probes) for comparison.

\begin{figure}
\begin{centering}
\includegraphics[angle=0,width=\linewidth]{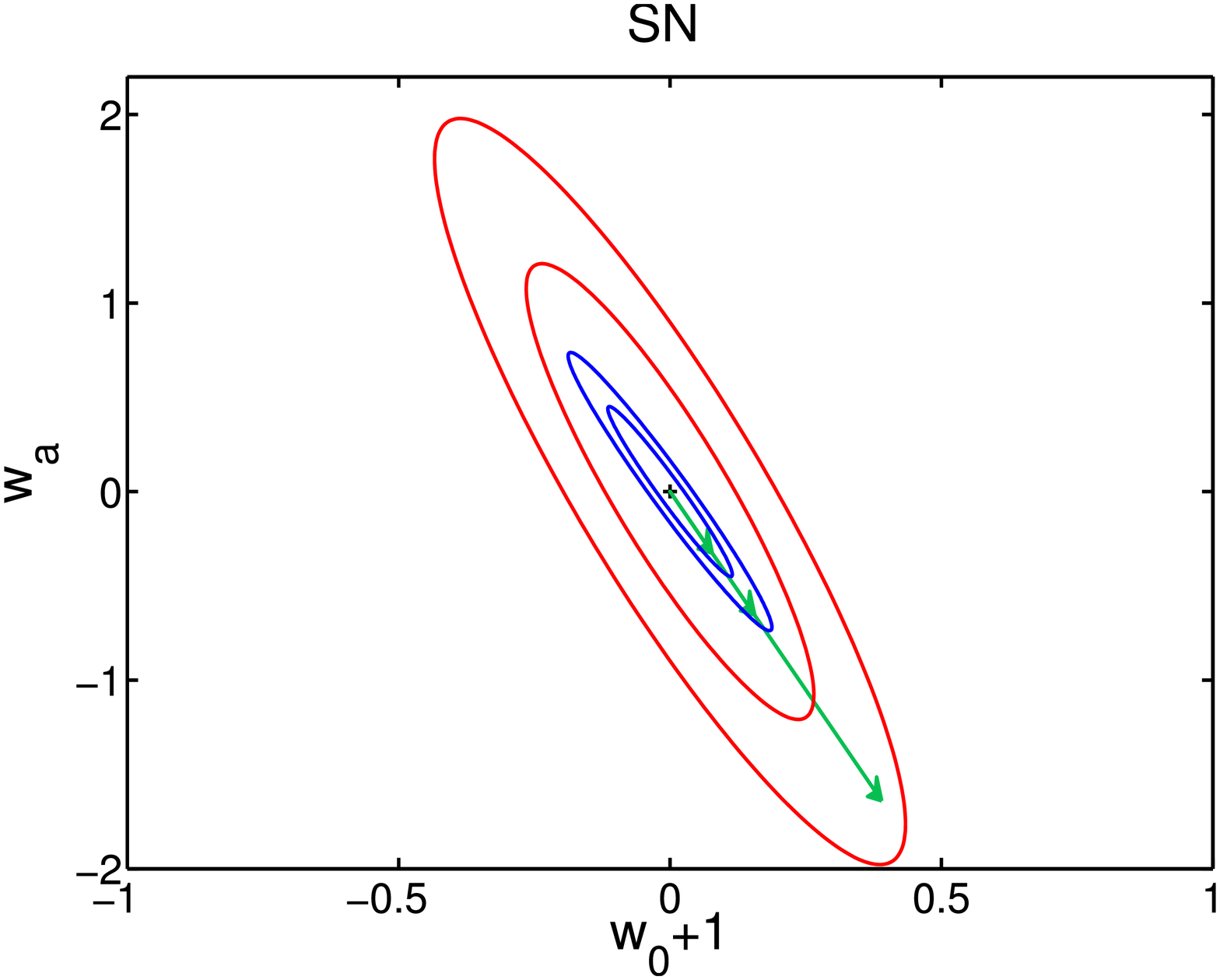} \\
\includegraphics[angle=0,width=\linewidth]{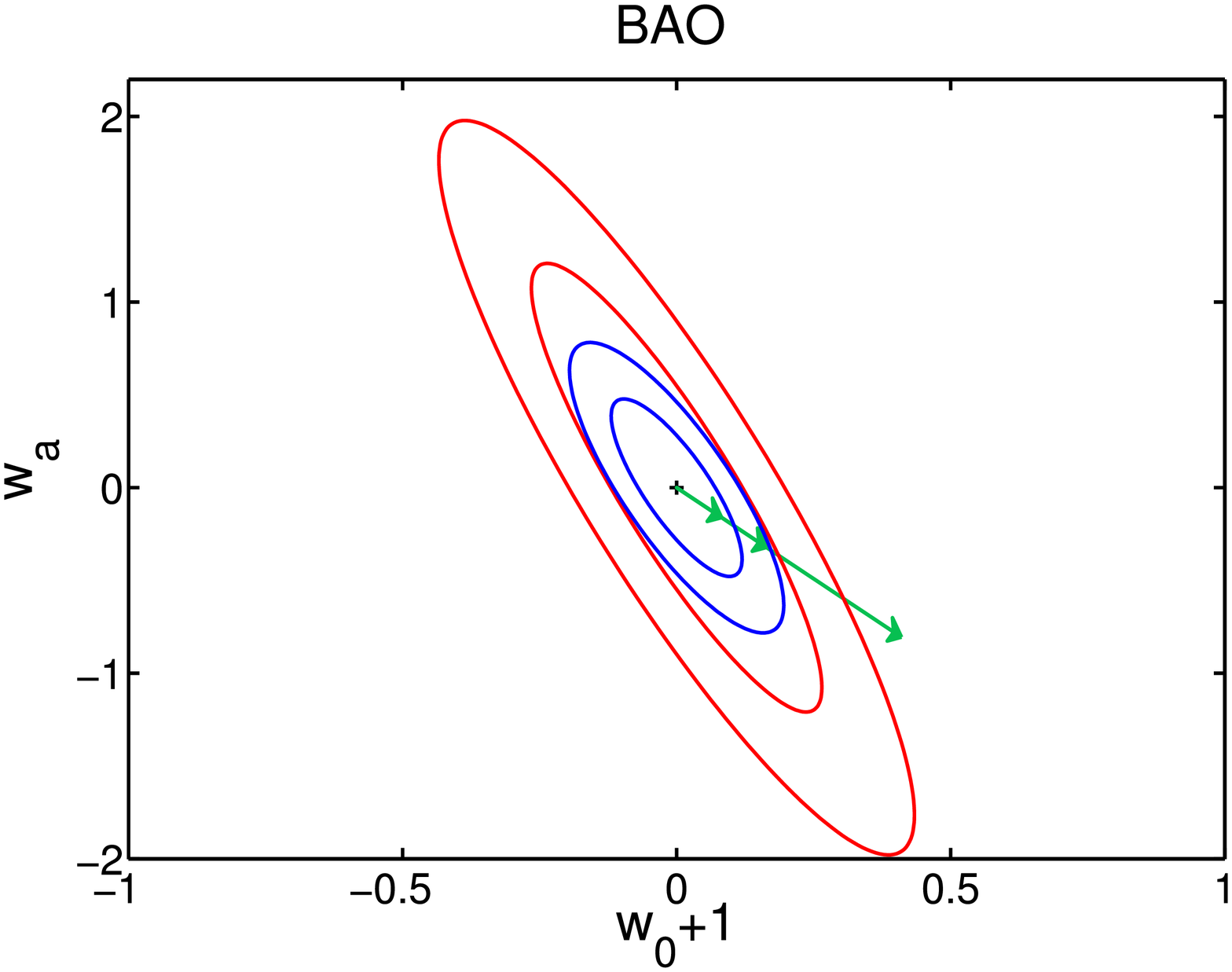} 
\par\end{centering}
\caption{Construction of the robustness FoM for a future SNIa survey (top
  panel) and a future BAO Euclid-like survey (bottom panel). Red ellipses show
  current 68\%, 95\% constraints (in a Gaussian approximation) from a
  combination of all available probes, blue ellipses show projected
  constraints from the future probe at the fiducial point (assumed to be
  $\Lambda$CDM). The green vectors show the systematic maximum excursion bias
  (MEB) for systematic erros of 1\%, 2\% and 5\% for SNIa and 0.1\%, 0.2\% and
  0.5\% for BAO. } \centering{}\label{fig:BAO_SN_ave}
\end{figure}

First, by inspecting Fig.~\ref{fig:BAO_SN_ave}, we notice that the systematic
bias projected onto the $(w_0, w_a)$ plane is much better aligned with the
degeneracy direction of the probes for SNIa than for BAO. From our discussion
in section~\ref{sec:properties}, this leads to expect a higher value for the
robustness FoM for SNIa than for BAO. Also, the size of the bias vectors in
the dark energy parameters is roughly comparable for SNIa and BAO, although in
the latter case we have adopted a bias in the observables ($H$ and $D$) which
is a factor of 10 smaller than for the SNIa observables (the
magnitudes). Table~\ref{tab:ave_results} shows that indeed both the robustness
and the average robustness FoMs are slightly larger for SNIa than for BAO
across the range of systematic error levels we adopted for each probe. This is
a consequence of the fact that the BAO bias leads to a smaller degree of
overlap of the BAO constraints with the present-day constraints, which is a
more serious lack of robustness than for the SNIa. In the latter case,
although the bias vectors are slightly larger in the dark energy parameters
(typically by a factor of 2, cf Table~\ref{tab:ave_results}), the bias
direction is well aligned with the statistical degeneracy, and therefore the
reduction in the overlap between the present constraints and future SNIa
constraints is less severe, translating in a higher robustness. For the
highest level of systematic error in each case (0.5\% for BAO and 5\% for
SNIa), we find that the robustness FoM for BAO is about a factor of 10 smaller
than for SNIa. The average robustness of BAO is also smaller, but only by
about 1/3, which reflects the more balanced assessment given by the average
robustness. Thus, for our particular choice of systematics, our findings run against the general lore that BAO
observations are more robust to systematics than SNIa.

In terms of our statistical FoM, the SNIa survey is better by a factor of
about 3, in good agreement with the result obtained from the usual DETF
FoM. Taken together, the better values of both the statistical and robustness
FoM for SNIa lead to a higher value of the total FoM for SNIa than for BAO.

It is important to stress that our robustness results above are {\em not} a
generic feature of SNIa and BAO observations. Rather, they reflect our
specific choices for the systematic bias in the observables for BAO and
SNIa. Other choices of systematic bias are possible and will in general give a
different results for the robustness, which we shall explore in a dedicated
paper.

%
%-----x-X- BEG FIGURE: SN AND BAO 28.11.2010 -X-x---------
%
%-----x-X- END FIGURE: SN AND BAO 28.11.2010 -X-x---------
%
%-----x-X- BEG AVE. RESULTS TABLE: SN AND BAO 22.11.2010 -X-x---------
\begin{table*} 
\centering
\begin{tabular}{lll | lll | lll }
\hline
             &&              &  \multicolumn{3}{c|}{BAO} &\multicolumn{3}{c}{SNIa}       \\   
              &&             & \multicolumn{3}{c|}{Maximum excursion bias}   &\multicolumn{3}{c}{Maximum excursion bias}      \\   
Fom & Symbol & Defined in                          &0.1\%  &0.2\%     &0.5\%                       & 1\%   &2\%  &5\%   \\ \hline  \hline
Robustness & $R_N$ & Eq.~\eqref{eq:R_N}        &$1.4$ &$0.83 $  &$0.026 $                      &$ 1.7$   &$ 1.3$ &$0.24$\\ 
Average robustness & $\avR$ &  Eq.~\eqref{eq:R_G}&1.4    &1.1       &0.54                        & 1.7     & 1.4   &0.81  \\ 
Statistical FoM & $S$ & Eq.~\eqref{eq:S}      &\multicolumn{3}{c|}{2.7}                       &\multicolumn{3}{c}{7.0} \\ 
Total FoM & $T_N$ & Eq.~\eqref{eq:T_N}          &3.6    &$2.2 $&$0.070$                         & 11.9    &9.1    &1.7 \\
Total average FoM & $\avT$ &  Eq.~\eqref{eq:Tav} &3.8    &2.9   &1.4                             & 11.9    &9.8    &5.7\\ 
Bias length & $|b|$ & caption  &0.18   &0.36   &0.90                            & 0.34    & 0.68  & 1.7     \\ 
\hline
DETF FoM & &  Eq.~\eqref{eq:DETF}   &\multicolumn{3}{c|}{4.4}     &\multicolumn{3}{c}{13}  \\

\end{tabular}
\caption{Robustness and statistical Figure of Merits for future BAO and SNIa
  surveys, for different levels of systematic errors in the observables.  We
  also give the DETF FoM for comparison (normalized to its value from current
  constraints). We also give the length of the bias vector $b$ in the $(w_0,
  w_a)$ plane. The Maximum excursion bias errors refer to both
    $D(z_b)$ and $H(z_b)$ in the case of BAO, and $m(z_b)$ in the case of SNIa.}
\label{tab:ave_results}
\end{table*}
%
%-----x-X- END AVE. RESULTS TABLE: SN AND BAO 22.11.2010 -X-x---------

\section{Conclusions}

\label{sec:conclusions}

We have introduced a new formalism to quantify the robustness of future dark
energy probes to systematic bias, and argued that this important new quantity
should be taken into account when evaluating the performance of future
surveys. In constrast to usual measures of statistical performance, our
robustness FoMs depend on the direction and size of the systematic bias
induced in the dark energy parameters by residual systematics in the
observables. We have thus described an approach to include the effect of systematic errors in the
  dark energy figures of merit.
%propagate systematic errors onto the dark energy parameter space.

We have applied this formalism to future SNIa and BAO probes by developping a
simple phenomenological model of possible residual systematic errors. Our
results indicate that -- for the specific choice of systematics
adopted here -- SNIa are slightly more robust to systematics than BAO, despite
having assumed a systematic shift in the observables for SNIa which is a
factor of 10 larger than for BAO. Coupled with the higher statistical
performance of SNIa, this would lead to prefer SNIa over BAO in terms of their
overall FoM. It is clear however that this particular result cannot be generalized
beyond our choice of systematics and surveys. In a future work we will investigate
how this result change by adopting more refined descriptions of the systematic bias for each
probe.

\bigskip
\textit{Acknowledgements.} L.A., D.H. and R.T. would like to thank the Galileo
Galilei Institute for Theoretical Physics for hospitality. M.C.M. would like
to thank the University of Heidelberg for hospitality. L.A. would like to
thank Imperial College London for hospitality. R.T. would like to thank the
INFN and the EU FP6 Marie Curie Research and Training Network ``UniverseNet''
(MRTN-CT-2006-035863) for partial support.  L.A. is partially supported by DFG
TRR33 ``The Dark Universe''. DH is supported by DOE OJI grant under contract
DE-FG02-95ER40899, NSF under contract AST-0807564, and NASA under contract
NNX09AC89G.

\section*{Appendix A}

The generalization of Eq.~\eqref{eq:FoM_2data} to an arbitrary number
of probes proceeds as follows. First, we notice that one can always
summarize current constraints from several observations in one single
joint posterior. Let us call the data from the combination of all
available present-day probes $d_{0}$ (with Fisher matrix $F_{0}$).
If one wishes to consider $N$ future probes, we can ask whether all
of the $N$ probes are mutually compatible%
\footnote{An alternative test would be to check whether the $N$-th probe is
compatible with the previous $N-1$ (assuming those are already available
and they are free of systematics themselves). In this case the relevant
quantity is 
\begin{equation}
R_{\text{N}}=\frac{p(d_{N}|d_{N-1}\dots d_{1})}{p(d_{N})p(d_{N-1}\dots d_{1})}\end{equation}
 which can be computed by appropriate substitutions in Eq.~\eqref{eq:FoM_2data}.%
}. Eq.~\eqref{eq:simplify_R} gives in this case \begin{eqnarray}
R_{\text{all}} & = & \frac{p(d_{N}d_{N-1}\dots d_{1}|d_{0})}{\prod_{j=1}^{N}p(d_{j}|d_{0})}\\[0.2cm]
 & = & \prod_{j=1}^{N}\frac{p(d_{j}|d_{j-1}\dots d_{1}d_{0})}{p(d_{j}|d_{0})}\\[0.2cm]
 & = & \prod_{j=2}^{N}\frac{p(d_{j}|d_{j-1}\dots
  d_{1}d_{0})}{p(d_{j}|d_{0})}\label{eq:Rall}
\end{eqnarray}
where in the last line we have cancelled out the very last term in
both the numerator and the denominator, so that the sum starts with
$j=2$. We now refer to Eq.~(\ref{eq:pd2d1}) to obtain \begin{eqnarray}
 &  & p(d_{j}|d_{j-1}\dots d_{1}d_{0})=\mathcal{L}_{0}^{(j)}\frac{|F_{012\dots(j-1)}|^{1/2}}{|F_{012\dots j}|^{1/2}}\label{eq:Rall_num}\\
 & \times & \exp\left[-\frac{1}{2}\left(\mu_{j}^{t}L_{j}\mu_{j}\right.\right.\nonumber \\
 & + & \mu_{012\dots(j-1)}^{t}F_{012\dots(j-1)}\mu_{012\dots(j-1)}\nonumber \\
 & - & \left.\left.\mu_{012\dots j}^{t}F_{012\dots j}\mu_{012\dots j}\right)\right],\nonumber \end{eqnarray}
 where the definitions correspond to those before, so that $F_{012\dots j}\equiv F_{0}+\sum_{i=1}^{j}L_{i}$,
and in particular $F_{012\dots N}\equiv F$. Notice already that most
terms in the numerator of Eq.~(\ref{eq:Rall}) will cancel.
Similarly, following Eq.~(\ref{eq:pd2}) \begin{eqnarray} \label{eq:Rall_denom}
p(d_{j}|d_{0}) & = & \mathcal{L}_{0}^{(2)}\frac{|F_{0}|^{1/2}}{|F_{0j}|^{1/2}}\\
 & \times & \exp\left[-\frac{1}{2}\left(\mu_{j}^{t}L_{j}\mu_{j}+\mu_{0}^{t}F_{0}\mu_{0}-\mu_{0j}^{t}F_{0j}\mu_{0j}\right)\right],\nonumber \end{eqnarray}
 Now one can evaluate Eq.~(\ref{eq:Rall}) with the help of Eqs.~(\ref{eq:Rall_num})
and (\ref{eq:Rall_denom}): 
 \begin{equation}
 \begin{split}
R_{\text{all}} & =  \frac{|F_{01}|^{1/2}}{|F|^{1/2}} \prod_{j=2}^{N}\frac{|F_{0}|^{-1/2}}{|F_{0j}|^{-1/2}}  \times  \exp\left[-\frac{1}{2}(\mu_{01}^{t}F_{01}\mu_{01}-\mu^{t}F\mu \right.\\
 & - \left.(N-1)\mu_{0}^{t}F_{0}\mu_{0}+\sum_{j=2}^{N}\mu_{0j}^{t}F_{0j}\mu_{0j})\right]
  \end{split}
 \end{equation}
 
 We thus obtain  for the robustness 
 \begin{equation}
 \begin{split}
\ln R_\text{all} & =  \frac{1}{2}\left(\sum_{i=1}^{N}\ln|F_{0i}|-(N-1)\ln|F_{0}|-\ln F\right)  \\
 & -  \frac{1}{2} ( \sum_{i=1}^{N}\overline{\mu}_{0i}^{t}F_{0i} \overline{\mu}_{0i}  -  (N-1) \mu_{0}^{t}F_{0}\mu_{0}-\mu^{t}F\mu ), 
 \end{split}
 \end{equation}
which generalizes Eq.~\eqref{eq:FoM_2data}.

%\bibliographystyle{mn2e}

%\bibliography{robust_biblio}

\end{document}